\documentclass[12pt]{article}
\usepackage{amsmath}
\begin{document}
\begin{flushright}
KANAZAWA-14-08\\
September, 2014
\end{flushright}
\vspace*{1cm}

\renewcommand\thefootnote{\fnsymbol{footnote}}
\begin{center} 
{\Large\bf Inflation in a modified radiative seesaw model}
\vspace*{1cm}

{\Large Romy H. S. Budhi}~$^{1,2,}$\footnote[1]{e-mail:
~romyhanang@hep.s.kanazawa-u.ac.jp}, ~
{\Large Shoichi Kashiwase}~$^{2,}$\footnote[2]{e-mail:
  ~shoichi@hep.s.kanazawa-u.ac.jp}, \\
{\Large and}\\
{\Large Daijiro Suematsu}~$^{2,}$\footnote[3]{e-mail:
~suematsu@hep.s.kanazawa-u.ac.jp}
\vspace*{0.5cm}\\

$^1${\it Physics Department, Gadjah Mada University, 
Yogyakarta 55281, Indonesia}\\
$^2${\it Institute for Theoretical Physics, Kanazawa University, 
Kanazawa 920-1192, Japan}
\end{center}
\vspace*{1.5cm} 

\noindent
{\Large\bf Abstract}\\
The existence of the inflationary era in the early Universe seems to
be strongly supported by recent CMB observations.
However, only a few realistic inflation scenarios which have 
close relation to particle physics seem to have been known unfortunately.  
The radiative neutrino mass model with inert doublet dark matter 
is a promising model for the present experimental issues which 
cannot be explained within the standard model.
In order to make the model include inflation, we extend it by a 
complex scalar field with a specific potential.
This scalar could be closely related to the neutrino mass generation 
at a TeV scale as well as inflation. We show that the inflation 
favored by the CMB observations could be realized even if inflaton 
takes sub-Planck values during inflation.               
\newpage
\setcounter{footnote}{0}
\renewcommand\thefootnote{\alph{footnote}}
\section{Introduction}
Recent discovery of a Higgs-like particle \cite{higgs} suggests 
that the framework of the standard model (SM) can describe Nature 
well up to the weak scale. On the other hand, we have experimental 
results which cannot be 
explained within it, that is, the existence of small neutrino masses 
\cite{nexp,t13}, the existence of dark matter \cite{uobs}, 
and baryon number asymmetry in the Universe \cite{baryon}. 
They require some extension of the SM. 

As such an example, we have a model which is the simple extension of the SM 
with a second doublet scalar (which has no vacuum expectation value and 
is called by several names such as inert \cite{inert}, 
scotogenic \cite{scot}, or doumant \cite{doum}) and also three 
right-handed neutrinos.
The model shows promising features in physics at TeV regions
for the explanation of both the neutrino oscillation data and the observed 
abundance of dark matter (DM).
In fact, if these new fields are assigned odd parity of an assumed $Z_2$
symmetry, small neutrino masses are generated at one-loop level and
the lightest $Z_2$ odd field can be stable as a DM candidate \cite{ma}.
The quantitative conditions required for their explanation 
in both this model and several extended models have been clarified through 
various studies by now \cite{f-nradm,ham,susy,inf-nradm,ks}.
They show that the simultaneous explanation of these is possible 
without causing a strong tension with other phenomena like lepton 
flavor violating processes if DM is identified with the lightest 
neutral component of the inert doublet scalar \cite{ham,ks}.
In such a case, moreover, the baryon number asymmetry in the Universe 
is also successfully explained if the resonant leptogenesis could occur
due to the mass degeneracy among right-handed neutrinos which have 
masses of a TeV scale \cite{resonant}. 
An interesting point is that the required 
mass degeneracy could be rather mild compared with the ordinary cases 
\cite{ks}.

The CMB observations suggest that the exponential
expansion of the Universe occurs in the very early Universe.
These results can constrain severely the allowed inflation models 
now \cite{planck,bicep2}.
For example, BICEP2 recently suggests that the tensor to scalar perturbation 
ratio should be $r\sim 0.2$ and the Hubble parameter
during inflation should take a value of $O(10^{14})$ GeV.
Although we know that a quadratic chaotic inflation model could 
be such a candidate, the inflaton should take trans-Planckian 
values during inflation in that model.
Since higher order terms which are suppressed by the Planck mass 
are generally expected to give larger contributions to the potential there, 
the flatness of potential cannot be guaranteed without any symmetry.

On the other hand, we do not have a lot of examples of inflaton 
that plays any role in particle physics. 
Inflaton is introduced just to bring about the inflation in many models.
As an exceptional example, one may suppose sneutrino 
inflation \cite{sneutrino}.\footnote{Higgs inflation \cite{infhiggs} and 
axionic inflation \cite{infaxion} are also motivated by particle physics.} 
If we consider the neutrino mass generation based on the seesaw 
mechanism in supersymmetric models, right-handed sneutrinos are introduced 
inevitably. One of them could work as inflaton causing the quadratic 
chaotic inflation. However, the model could be annoyed by the above 
mentioned trans-Planckian problem. 

In this paper, we propose an inflation scenario in the framework 
related to the radiative seesaw model. 
Although several inflation scenarios have been considered 
in the radiative seesaw model, they have 
still problems, for example, the above mentioned trans-Planckian 
problem \cite{inf-nradm} or the unitarity problem caused by a large 
non-minimal coupling \cite{rad-infl}. 
Our scenario is based on an extension of the radiative seesaw model 
with a complex scalar, whose component is identified with the inflaton. 
We show that sufficient e-foldings could be realized even if the 
inflaton takes sub-Planckian values during inflation. 
In this scenario, the scalar spectral index and the tensor-to-scalar 
ratio could have values in the region favorable from the recent 
precise CMB observations.
In particular, the tensor-to-scalar ratio could take rather wide 
range values consistent with the CMB results depending on the 
parameters in the inflaton potential.
Moreover, this inflaton could play a crucial role for the neutrino 
mass generation other than the inflation, which is similar to 
the sneutrino inflation scenario.  

The paper is organized as follows. In the next section we address our
extended model briefly. In particular, the role of a new singlet scalar 
in the neutrino mass generation is explained. 
In section 3, we study the inflation in this model. Important quantities
relevant to the inflation such as e-foldings, slow-roll parameters and
spectral index are estimated numerically.
Reheating temperature is also discussed. 
In section 4 we summarize the paper.

\section{An extended model}
The original radiative seesaw model is defined by the following $Z_2$ 
invariant terms \cite{ma}: 
\begin{eqnarray}
-{\cal L}_O&=&-h_{\alpha i} \bar N_i\eta^\dagger\ell_\alpha
-h_{\alpha i}^\ast\bar\ell_\alpha\eta N_i+
\frac{M_i}{2}\bar N_iN_i^c 
+\frac{M_i^\ast}{2}\bar N_i^cN_i \nonumber \\
&+&m_\phi^2\phi^\dagger\phi+m_\eta^2\eta^\dagger\eta+
\lambda_1(\phi^\dagger\phi)^2+\lambda_2(\eta^\dagger\eta)^2
+\lambda_3(\phi^\dagger\phi)(\eta^\dagger\eta)  
+\lambda_4(\eta^\dagger\phi)(\phi^\dagger\eta)  \nonumber\\
&+&\frac{1}{2}\left[\lambda_5(\eta^\dagger\phi)^2 
+\lambda_5^\ast(\phi^\dagger\eta)^2 \right],
\label{model1}
\end{eqnarray}
where $\ell_i$ is a left-handed lepton doublet and $\eta$ 
is an inert doublet scalar. 
Since $\eta$ and right-handed neutrinos $N_i$ 
are assigned odd parity of $Z_2$ symmetry and 
all the SM contents including the ordinary Higgs doublet scalar $\phi$ 
are assigned even parity,  
neutrino Dirac mass terms are forbidden at tree-level. 
Neutrino masses are generated through one-loop diagram which 
has $N_i$ and $\eta$ in the internal lines as shown in the left 
diagram of Fig.~1. 

As found from this figure, neutrino mass generation 
in this model is characterized by a scalar quartic coupling
$\lambda_5(\eta^\dagger\phi)^2$ between the ordinary doublet Higgs 
scalar $\phi$ and an inert doublet scalar $\eta$. 
In this mass generation at TeV scales, the smallness of the 
coupling $\lambda_5$ plays a crucial role to explain the smallness 
of neutrino masses. It is considered to be a key feature of this scenario.
To explain its smallness, we may consider a scenario that this coupling 
is an effective coupling appearing at low energy regions after 
integrating out a heavy complex singlet scalar $S$.
Such a scenario could be realized by introducing a $Z_2$ odd 
singlet complex scalar.
Additional terms in the new Lagrangian are given as  
\begin{eqnarray}
-{\cal L}_S&=& \tilde m_S^2S^\dagger S + \frac{1}{2}m_S^2S^2 
+\frac{1}{2} m_S^2S^{\dagger 2}+
\kappa_1(S^\dagger S)^2 
+\kappa_2(S^\dagger S)(\phi^\dagger\phi)+ \kappa_3(S^\dagger S)(\eta^\dagger\eta) 
\nonumber \\
&-& \mu S\eta^\dagger\phi - \mu^\ast S^\dagger\phi^\dagger\eta, 
\label{model}
\end{eqnarray}
where these are most general terms which are $Z_2$ invariant 
and renormalizable.\footnote{In this extension, $\lambda_5=0$ is 
supposed in eq.~(\ref{model1}). Since the $\beta$-function of $\lambda_5$
is proportional to $\lambda_5$ itself, this assumption is justified after 
taking into account the radiative correction.}

\input epsf
\begin{figure}[t]
\begin{center}
\epsfxsize=12cm
\leavevmode
\epsfbox{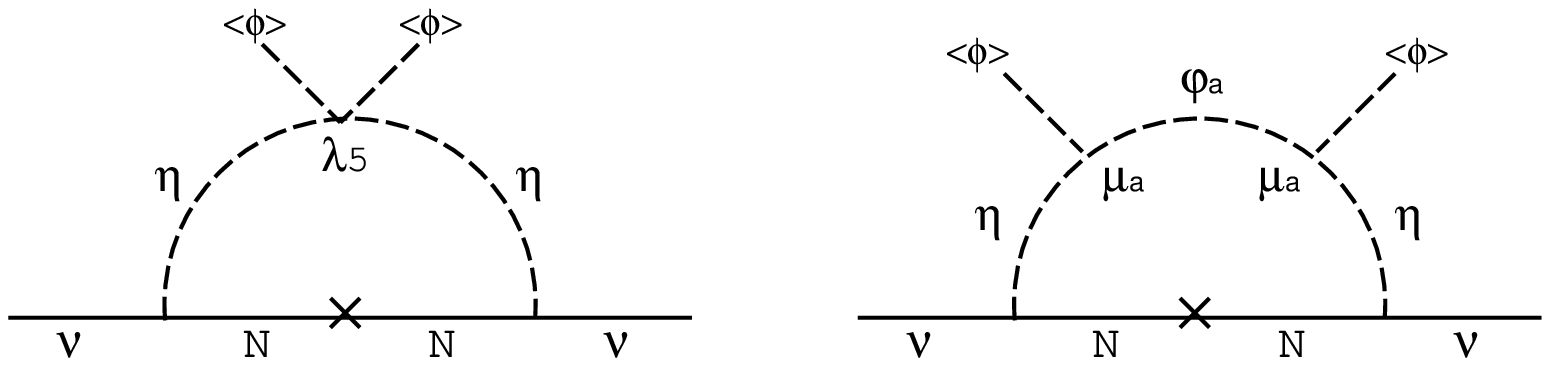}
\end{center}
\vspace*{-3mm}

{\footnotesize {\bf Fig.~1}~~One-loop diagrams which contribute 
neutrino mass generation. Left one is the diagram in the original 
model. The right diagram generates neutrino masses in the present model.
The dimensionful coupling $\mu_a$ is defined as 
$\mu_1=\frac{\mu}{\sqrt 2}$ and $\mu_2=\frac{i\mu}{\sqrt 2}$ by 
using $\mu$ in eq.~(\ref{model})}.  
\end{figure}

If the new singlet $S$ is much heavier than $\eta$ and $N_i$, 
favorable features of the original model could be kept in this extended 
version. 
Neutrino mass is generated through the one-loop diagram shown in the right 
one of Fig.~1. In this diagram, $\varphi_a$ represents the component 
fields of $S$ which are defined as $S=\frac{1}{\sqrt 2}(\varphi_1+i\varphi_2)$.
Their masses are easily found to be $\bar m_1^2=\tilde m_S^2+m_S^2$ and 
$\bar m_2^2=\tilde m_S^2-m_S^2$. 
Since $Z_2$ is considered as an exact symmetry, $\tilde m_S^2 >m_S^2$ should 
be satisfied.
The similar diagram is know to contribute to the 
neutrino mass generation in the supersymmetrized radiative 
seesaw model \cite{susy}.
Neutrino mass induced through this diagram can be estimated as
\begin{equation}
({\cal M}_\nu)_{\alpha\beta}=\sum_{i=1}^3\sum_{a=1,2}
\frac{h_{\alpha i}h_{\beta i}M_i\mu_a^2\langle\phi\rangle^2}{8\pi^2}
I(M_{\eta}, M_i, \bar m_a), 
\label{nmtr2}
\end{equation}
where $M_\eta^2=m_\eta^2+(\lambda_3+\lambda_4)\langle\phi\rangle^2$ and
$\mu_a$ is defined as $\mu_1=\frac{\mu}{\sqrt 2}$ and 
$\mu_2=\frac{i\mu}{\sqrt 2}$, respectively.
The function $I(m_a,m_b,m_c)$ is defined as
\begin{eqnarray}
I(m_a,m_b,m_c)&=&\frac{(m_a^4-m_b^2m_c^2)~\ln m_a^2}
{(m_b^2-m_a^2)^2(m_c^2-m_a^2)^2}+
\frac{m_b^2~\ln m_b^2}
{(m_c^2-m_b^2)(m_a^2-m_b^2)^2} \nonumber\\
&+&\frac{m_c^2~\ln m_c^2}
{(m_b^2-m_c^2)(m_a^2-m_c^2)^2}-
\frac{1}{(m_b^2-m_a^2)(m_c^2-m_a^2)}.
 \label{mnu2}
\end{eqnarray} 
If we assume that the conditions $\tilde m_S \gg m_S, m_\eta, M_i$ are 
satisfied, this formula can be approximated as
\begin{equation}
({\cal M}_\nu)_{\alpha\beta}=\sum_{i=1}^3
\frac{h_{\alpha i}h_{\beta i}\langle\phi\rangle^2}{8\pi^2}
\frac{m_S^2\mu^2}{\tilde m_S^4}
\frac{M_i}{M_\eta^2-M_i^2}\left[
\frac{M_i^2}{M_\eta^2-M_i^2}\ln \frac{M_i^2}{M_\eta^2}+1\right].
\end{equation}
It is equivalent to the neutrino mass formula in the original model if
$\frac{m_S^2\mu^2}{\tilde m_S^4}$ is identified with the coupling constant 
$\lambda_5$ for the $(\eta^\dagger\phi)^2$ term.

This correspondence might be found in an effective theory obtained at 
energy regions smaller than $\tilde m_S$ by integrating out $S$. 
In fact, if we use the equation of motion for $S$ which could be 
approximated as $S\simeq \mu^\ast\phi^\dagger\eta/\tilde m_S^2$,
the required terms are derived as
\begin{equation}
\frac{1}{2}\left[\frac{m_S^2\mu^2}{\tilde m_S^4}(\eta^\dagger\phi)^2
+ \frac{m_S^2\mu^{\ast 2}}{\tilde m_S^4}(\phi^\dagger\eta)^2\right].
\label{model2}
\end{equation}
The origin of small $\lambda_5$ which is the key nature to explain the 
smallness of the neutrino masses is now translated to the hierarchy 
problem between $\mu$, $m_S$ and $\tilde m_S$ in this extension.
If we leave the origin of this hierarchy to a complete theory 
at high energy regions, all the neutrino masses, the DM abundance 
and the baryon number asymmetry could be also explained in this extended 
model at TeV regions just as in the same way discussed in the 
previous articles \cite{ks}. 
Following the results obtained in these studies, the value of 
$\frac{m_S^2\mu^2}{\tilde m_S^2}$ could be constrained by the simultaneous 
explanation of these.

\section{Inflation due to the complex scalar $S$}
\subsection{e-foldings and the spectral index}
If the singlet scalar $S$ does not play any other role, 
this modification might not be so interesting. However, we find that 
the introduction of $S$ could add an interesting feature to the 
radiative seesaw model as an inflation model.\footnote{Higgs inflation 
has been applied to the radiative seesaw model in \cite{rad-infl}.}
As such simple scenarios for a real singlet scalar $S$, one may consider 
$m_S^2S^2$ type chaotic inflation \cite{inf-nradm} or $S$-inflation 
\cite{sinfl}. In the former example, the inflation could be related 
with the neutrino mass generation like sneutrino inflaton model.
However, the scenario requires trans-Planckian values for $S$
during inflation and it could induce the above mentioned problem.

In this section, we consider an inflation scenario which could work even 
for sub-Planckian values of $S$, following the proposal in \cite{mac}. 
We show that it is possible as long as the existence of specific 
nonrenormalizable terms is assumed in the potential for $S$.
As such potential, we suppose that the complex scalar $S$ has $Z_2$ 
invariant additional potential terms such as 
\begin{eqnarray}
V&=& c_1\frac{(S^\dagger S)^n}{M_{\rm pl}^{2n-4}}
\left[1+ c_2\left\{ \left(\frac{S}{M_{\rm pl}}\right)^{2m} 
\exp\left(i\frac{S^\dagger S}{\Lambda^2}\right)
+ \left(\frac{S^\dagger}{M_{\rm pl}}\right)^{2m}
\exp\left(-i\frac{S^\dagger S}{\Lambda^2}\right)
\right\} \right] \nonumber \\
&=&c_1\frac{\varphi^{2n}}{2^nM_{\rm pl}^{2n-4}}\left[
1+ 2c_2\left(\frac{\varphi}{\sqrt 2 M_{\rm pl}}\right)^{2m}
\cos\left(\frac{\varphi^2}{2\Lambda^2}+2m\theta\right)\right],  
\label{model3}
\end{eqnarray}
where both $n$ and $m$ are positive integers 
and $M_{\rm pl}$ is the reduced Planck mass.
We use the polar coordinate expression $S=\frac{\varphi}{\sqrt 2}
 e^{i\theta}$ in the second equality of eq.~(\ref{model3}). 
Although the exponential part of this potential is crucial for 
the present scenario, we cannot describe its origin concretely 
at the present stages. We only expect that it might be effectively 
induced through the nonperturbative dynamics in the UV completion 
of the model. However, since the model has interesting features 
for the inflation as shown below, we expect that the potential 
form might give us some useful hints to find the UV completion 
and its dynamics. 

In the left panel of Fig.~2, we show a typical shape of 
the potential as a function of $\varphi$ for a fixed $\theta$.
As easily found, the potential $V$ has local 
minimums for a fixed $\theta$ under the condition 
$\Lambda \ll \varphi \ll M_{\rm pl}$ at 
\begin{equation}
\frac{\varphi^2}{2\Lambda^2}+2m\theta=(2j+1)\pi + \alpha
\simeq (2j+1)\pi,
\label{cinf}
\end{equation}
where $j$ is an integer and
$\alpha=\tan^{-1}\left(\frac{2\Lambda^2}{M_{\rm pl}^2}\right)$. 

\begin{figure}[t]
\begin{center}
\epsfxsize=7.5cm
\leavevmode
\epsfbox{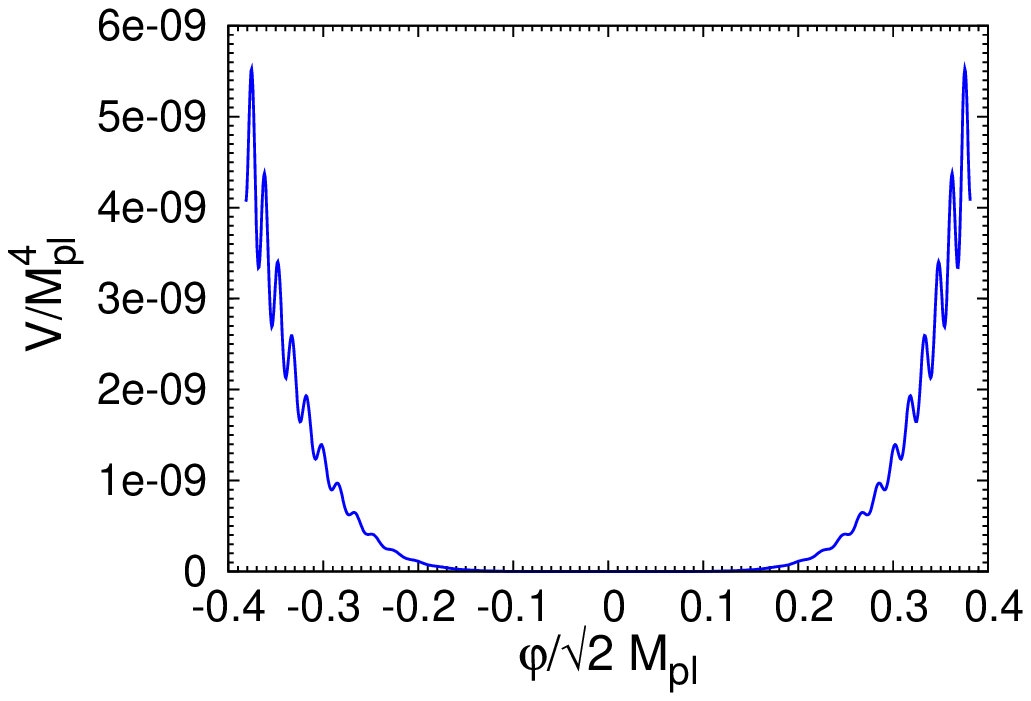}
\hspace{7mm}
\epsfxsize=7.5cm
\leavevmode
\epsfbox{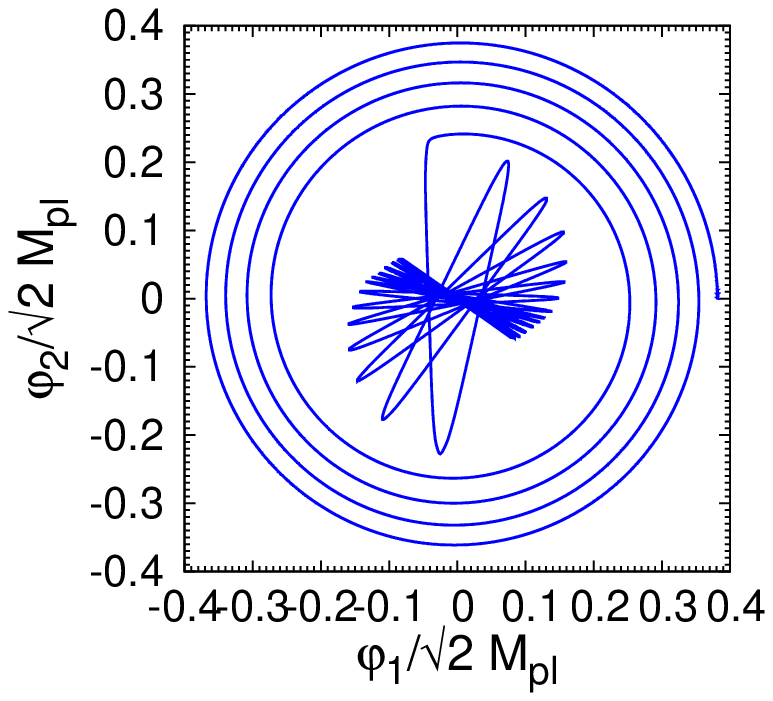}
\end{center}
\vspace*{-3mm}

{\footnotesize {\bf Fig.~2}~~The left panel shows the potential $V$ defined by
$n=3$ and $m=1$. Other parameters in $V$ are fixed as $c_1=1.65\times 10^{-6}$, 
$c_2=0.7$ and $\Lambda/M_{\rm pl}=0.04$. In the right panel, 
the time evolution of the field $a$ in the 
$(\frac{\varphi_1}{\sqrt 2M_{\rm pl}},\frac{\varphi_2}{\sqrt 2M_{\rm pl}})$ plane 
for the potential $V$ shown in the left panel. $\varphi$ is related to
$\varphi_{1,2}$ by $\varphi^2=\varphi_1^2+\varphi_2^2$.  }  
\end{figure}

Now we assume that inflation proceeds along this local minimum.
In that case, the field $a$ along this direction is considered to play 
a role of inflaton. It might be represented as 
$da^2=d\varphi^2+\varphi^2 d\theta^2$.
Since $\varphi$ is supposed to evolve as a function of $\theta$ 
following eq.~(\ref{cinf}), we find that the field $a$ should 
satisfy the relation such as 
\begin{equation}
da=\left[\varphi^2+\left(\frac{d\varphi}{d\theta}\right)^2\right]^{1/2}d\theta
=\left[1 + 4m^2\left(\frac{\Lambda}{\varphi}\right)^4\right]^{1/2}
\varphi d\theta.
\label{infl}
\end{equation}
This shows that the field $a$ can be expressed as $da\simeq\varphi d\theta$ 
as long as $\varphi \gg \Lambda$ is satisfied.
Thus, the field $a$ associated to the almost flat direction 
can be treated as a canonically normalized inflaton field orthogonal 
to $\varphi$.
In order to estimate the mass of $\varphi$ during the period when the 
field evolve along the $a$ direction, we expand the potential 
$V$ given in eq.~(\ref{model3}) around its local minimum at a fixed $\theta$. 
As this result, we find that the mass of $\varphi$ satisfies
\begin{equation} 
m_\varphi~{^>_\sim}~ \left[\frac{c_1}{2^{n-1}}n(2n-1)\right]^{1/2}
\left(\frac{\varphi}{M_{\rm pl}}\right)^{n-2}\varphi.
\end{equation}
On the other hand, the Hubble parameter during this period could be
roughly estimated as 
\begin{equation}
H=\left(\frac{V}{3M_{\rm pl}^2}\right)^{1/2}\simeq 
\left(\frac{c_1}{3\cdot 2^n}\right)^{1/2}
\left(\frac{\varphi}{M_{\rm pl}}\right)^{n-1}\varphi.
\end{equation}
As long as we suppose a situation such as $\varphi \ll M_{\rm pl}$,
we find that the mass of $\varphi$ is much larger than the Hubble 
parameter there.
This shows that $\varphi$ cannot contribute to the inflation and then  
the single inflaton scenario due to the field $a$ could be realized.

We can check that this actually occurs through numerical calculation. 
We solve the field equations for the component fields $\varphi_{1,2}$ 
of $S=\frac{1}{\sqrt 2}(\varphi_1+i\varphi_2)$ numerically. 
They evolve following the field equations,
\begin{equation}
\ddot{\varphi_i}+3H\dot{\varphi_i}=-\frac{\partial V}{\partial\varphi_i} 
\qquad (i=1,~2)
\label{evol}
\end{equation}
where $H^2=\frac{1}{3M_{\rm pl}^2}(\sum_i\frac{1}{2}\dot{\varphi}_i^2+V)$.
For example, the explicit expression for $\frac{\partial V}{\partial\varphi_i}$ 
in the case of $n=3$ and $m=1$ is given as
\begin{eqnarray}
\frac{\partial V}{\partial\varphi_1}&=& 
\frac{c_1(\varphi_1^2+\varphi_2^2)^3}{8M_{\rm pl}^2}
\left[\frac{6\varphi_1}{\varphi_1^2+\varphi_2^2}
+\frac{c_2\varphi_1}{M_{\rm pl}^2}
\left\{\frac{6(\varphi_1^2-\varphi_2^2)}{\varphi_1^2+\varphi_2^2}+2
-2\frac{\varphi_1\varphi_2}{\Lambda^2}\right\}
\cos\left(\frac{\varphi_1^2+\varphi_2^2}{2\Lambda^2}\right) \right.\nonumber \\
&-&\left. c_2\frac{\varphi_1}{M_{\rm pl}^2} \left\{
\frac{12\varphi_1\varphi_2}{\varphi_1^2+\varphi_2^2}
+2\frac{\varphi_2}{\varphi_1}
+\frac{\varphi_1^2-\varphi_2^2}{\Lambda^2}\right\}
\sin\left(\frac{\varphi_1^2+\varphi_2^2}{2\Lambda^2}\right)\right],
 \nonumber\\
\frac{\partial V}{\partial\varphi_2}&=&
\frac{c_1(\varphi_1^2+\varphi_2^2)^3}{8M_{\rm pl}^2}
\left[\frac{6\varphi_2}{\varphi_1^2+\varphi_2^2}
+\frac{c_2\varphi_2}{M_{\rm pl}^2}
\left\{\frac{6(\varphi_1^2-\varphi_2^2)}{\varphi_1^2+\varphi_2^2}-2
-2\frac{\varphi_1\varphi_2}{\Lambda^2}\right\}
\cos\left(\frac{\varphi_1^2+\varphi_2^2}{2\Lambda^2}\right) \right.\nonumber \\
&-&\left. c_2\frac{\varphi_2}{M_{\rm pl}^2} \left\{
\frac{12\varphi_1\varphi_2}{\varphi_1^2+\varphi_2^2}
+2\frac{\varphi_1}{\varphi_2}
+\frac{\varphi_1^2-\varphi_2^2}{\Lambda^2}\right\}
\sin\left(\frac{\varphi_1^2+\varphi_2^2}{2\Lambda^2}\right)\right].
\end{eqnarray} 
In the right panel of Fig.~2, we show an example for the evolution of 
the inflaton in the 
$(\frac{\varphi_1}{\sqrt 2M_{\rm pl}}, \frac{\varphi_2}{\sqrt 2M_{\rm pl}})$ 
plane. In this calculation, 
we assume that $\varphi_{1,2}$ initially stay at the local minimum.
The figure shows that the field $a$ evolves aperiodic circle.
Along this trajectory, the value of $a$ changes by an amount larger
than the Lyth bound \cite{lyth} during the small change of $\varphi$
in the sub-Planckian range. 
>From this figure, we find that the single inflaton scenario could 
be realized in this model as long as the conditions mentioned above 
are satisfied and the fields $\varphi_{1,2}$ start to evolve 
from a local minimum.

Next, in order to see the feature of the inflation induced by this field $a$,
we calculate the quantities which characterize inflation, that is, 
the e-foldings $N$, the spectral index $n_s$, the tensor-to-scalar ratio 
$r$ and so on. 
The change of the inflaton $a$ from some period to the end of inflation can be 
expressed by using $\varphi$ as
\begin{equation}
a_e-a=-\int^{\varphi_e}_\varphi \frac{\tilde\varphi^2}{2m\Lambda^2}
d\tilde\varphi=\frac{1}{6m\Lambda^2}(\varphi^3-\varphi_e^3),
\end{equation}
where we use eq.~(\ref{infl}) under the assumption $\varphi\gg \Lambda$. 
A value of $\varphi$ at the end of inflation is expressed by $\varphi_e$.
For the convenience, we may redefine the canonically normalized 
new inflaton as
\begin{equation}
\chi\equiv a_e+\frac{\varphi_e^3}{6m\Lambda^2}-a=
\frac{\varphi^3}{6m\Lambda^2}.
\end{equation}
This expression explicitly shows that Sub-Planckian values of $\varphi$
could be enhanced by $\frac{\varphi^2}{6m\Lambda^2}$ to result in 
trans-Planckian values of the inflaton $\chi$.  
The e-foldings induced by the inflaton change from $\chi$ to $\chi_e$ 
is estimated as
\begin{eqnarray}
N&=&-\frac{1}{M_{\rm pl}^2}\int_\chi^{\chi_e} d\chi ~\frac{V}{V^\prime}
\equiv N(\chi)-N(\chi_e), \nonumber \\
N(\chi)&=&\frac{1}{6m^2n}
\left(\frac{M_{\rm pl}}{\Lambda}\right)^4
\left(\frac{\varphi}{\sqrt 2M_{\rm pl}}\right)^6\left[~1 
+\frac{6c_2m}{n(3+m)}\left(\frac{\varphi}{\sqrt 2M_{\rm pl}}\right)^{2m}
\right. \nonumber \\
&&\hspace*{2cm}\left. 
\times F\left(1,~\frac{3}{m}+1,~\frac{3}{m}+2,~2c_2\left(1+\frac{m}{n}\right)
\left(\frac{\varphi}{\sqrt 2M_{\rm pl}}\right)^{2m}\right)\right],
\label{efold}
\end{eqnarray}
where $V^\prime=\frac{dV}{d\chi}$ and $F$ is the hypergeometric function.
$\chi_e$ is fixed as a value at the end of inflation. In the expression 
of $N(\chi)$, 
it might be approximated by the first term since the second term is 
negligibly small compared with it.
However, it should be noted that $N(\chi)\gg N(\chi_e)$ is not satisfied 
in this scenario. 

Slow-roll parameters \cite{slowroll} are easily calculated by 
using eqs.(\ref{model3}) and 
(\ref{infl}). We find that they are given by using the model parameters as
\begin{eqnarray}
\varepsilon&\equiv&\frac{M_{\rm pl}^2}{2}\left(\frac{V^\prime}{V}\right)^2
=m^2\left(\frac{\sqrt 2M_{\rm pl}}{\varphi}\right)^6
\left(\frac{\Lambda}{M_{\rm pl}}\right)^4\left[
\frac{n-2c_2(m+n)\left(\frac{\varphi}{\sqrt 2M_{\rm pl}}\right)^{2m}}
{1-2c_2\left(\frac{\varphi}{\sqrt 2M_{\rm pl}}\right)^{2m}}\right]^2, 
\nonumber \\
\eta&\equiv& M_{\rm pl}^2\left(\frac{V^{\prime\prime}}{V}\right) \nonumber\\
&=&m^2\left(\frac{\sqrt 2M_{\rm pl}}{\varphi}\right)^6
\left(\frac{\Lambda}{M_{\rm pl}}\right)^4
\frac{n(2n-3)-2c_2(m+n)(2m+2n-3)\left(\frac{\varphi}{\sqrt 2M_{\rm pl}}
\right)^{2m}}
{1-2c_2\left(\frac{\varphi}{\sqrt 2M_{\rm pl}}\right)^{2m}}. \nonumber\\  
\label{slow}
\end{eqnarray}
If $c_2$ terms are neglected in these formulas, we find that 
these slow-roll parameters at the period characterized by the 
inflaton value $\chi$ can be represented in the very 
simple forms such as $\varepsilon\simeq\frac{n}{6(N+N(\chi_e))}$ and 
$\eta\simeq\frac{2n-3}{6(N+N(\chi_e))}$ by using the e-foldings $N$ 
given in eq.~(\ref{efold}). We note that the explicit 
$m$ dependence in these quantities remains only in the expression 
of the e-foldings $N$. The end of inflation could occur much before
the time when $\varepsilon=1$ is realized. 
It is crucial to guarantee the field evolution along the local potential 
minimum and the $c_2$ term plays a key role there.

We clarify this feature through a simple observation.
When the kinetic energy is equal to the local potential barrier $V_b$ 
which is given by the $\cos$ term of eq.~(\ref{model3}),
the inflaton could go over the potential barrier to the global minimum.
The condition could be expressed as $\frac{1}{2}\dot{\chi}^2\sim V^\prime$.
If we use the slow-roll approximation $3H\dot{\chi}=-V^\prime$ here,
this condition can be written as $\varepsilon\sim \frac{3V_b}{V}$.
Since $V>V_b$ is supposed in the model, the end of inflation occurs 
at the time when $\varepsilon <1$ is satisfied. 
By solving this condition, we can estimate the value of $\varphi_e$ as
\begin{equation}
\frac{\varphi_e}{\sqrt 2M_{\rm pl}}\simeq 
\left(\frac{m^2n}{6c_2}\right)^{\frac{1}{2m+6}}
\left(\frac{\Lambda}{M_{\rm pl}}\right)^{\frac{2}{m+3}},
\label{phie}
\end{equation}
where we neglect the contribution from the $c_2$ terms.
In that case, $N(\chi_e)$ has a substantial contribution to determine 
the e-foldings $N$.
To confirm this behavior and estimate the value of $\chi_e$, 
we use the numerical solutions of the field equations (\ref{evol}) 
which contain the effect of the $c_2$ term.
The numerical results given in the latter part show a good agreement
with the values of $\chi_e$ derived by using eq.~(\ref{phie}).
It supports our picture for the end of inflation.  
  
The spectrum of scalar perturbation predicted by the inflation 
is expressed as \cite{slowroll}
\begin{equation}
{\cal P_R}(k)=\Delta_{\cal R}^2\left(\frac{k}{k^\ast}\right)^{n_s-1},  \qquad
\Delta_{\cal R}^2=\frac{V}{24\pi^2M_{\rm pl}^4\varepsilon}\Big|_{k^\ast}. 
\label{power}
\end{equation}
The CMB observations give the normalization such that 
$\Delta_{\cal R}^2\simeq 2.43\times 10^{-9}$ at $k_\ast=0.002~{\rm Mpc}^{-1}$. 
This constrains the value of $V/\varepsilon$ at the time when 
the scale characterized by the wave number $k_\ast$ exits the horizon.
On the other hand, the remaining e-foldings $N_\ast$ of the inflation 
after the scale $k_\ast$ exits the horizon is dependent on the reheating 
phenomena and others as \cite{slowroll}
\begin{equation}
N_\ast\simeq 61.4-\ln\frac{k_\ast}{a_0H_0}
-\ln\frac{10^{16}~{\rm GeV}}{V_{k_\ast}^{1/4}}
+\ln\frac{V_{k_\ast}^{1/4}}{V_{\rm end}^{1/4}}-
\frac{1}{3}\ln\frac{V_{\rm end}^{1/4}}{\rho_{\rm reh}^{1/4}}.
\end{equation} 
Taking account of this uncertainty, $N_\ast$ is usually considered to 
take a value in the range 50 - 60. 
Here we also use the values in this range and we represent 
a value of $\varphi$ which gives the e-foldings $N_\ast$ as $\varphi_\ast$. 
If we use these notations, the above normalization $\Delta_{\cal R}^2$ is found 
to have a suitable value for
\begin{equation}
c_1=9.5\times 10^{-8}\frac{n}{N_\ast}
\left(\frac{\sqrt{2} M_{\rm pl}}{\varphi_\ast}\right)^{2n}, 
\label{c1}
\end{equation}
as long as the $c_2$ term in the potential is neglected.\footnote{One 
might find that this condition could be
easily satisfied even for a large value of $c_1$ near $O(1)$.
In fact, if we scale $c_1$, $c_2$ and $\Lambda$ such as $x^{2n}c_1$, 
$x^{2m}c_2$ and $x^{-1}\Lambda$ with $x$, the potential keeps 
its form for the scaled $x^{-1}\varphi$. Although the numerical detail 
in the field evolution has subtle behavior, the basic feature is understood 
in this way as found in the solution given in Table 1.} 
As examples, if we suppose $n=3$ and $N_\ast=60$,
$c_1\simeq 3\times 10^{-7}$ and 0.3 are required for 
$\frac{\varphi_\ast}{\sqrt 2}=0.5M_{\rm pl}$ and $0.05M_{\rm pl}$, 
respectively.

The scalar spectrum index $n_s$ and the ratio of the 
tensor perturbation to the scalar perturbation $r$ can be represented 
by using the slow-roll parameters $\varepsilon$ and $\eta$ as follows
\cite{slowroll},
\begin{equation}
n_s=1-6\varepsilon+2\eta, \qquad  r=16\varepsilon.
\end{equation}
If we use the formulas (\ref{slow}), 
we can estimate $n_s$ and $r$ at $k_\ast$ in this model. 
In particular, when $c_2$ terms are negligibly small, 
these are summarized by using the e-foldings $N$ in the very simple 
forms such as
\begin{equation}
n_s=1-\frac{n+3}{3(N_\ast+N(\chi_e))}, \qquad  
r=\frac{8n}{3(N_\ast+N(\chi_e))}.
\label{chaotic}
\end{equation} 
It is very interesting that both expressions of $n_s$ and $r$ 
given in eq.~(\ref{chaotic}) reduce to the same forms 
which are obtained in the chaotic inflation with the quadratic 
potential in the case $n=3$ as suggested in \cite{mac}.
They are known to be favored for reasonable $N_\ast$ values such 
as $N_\ast =50$ - 60 by BICEP2 results.
However, $N(\chi_e)\simeq 0$ is not guaranteed in the present model 
as mentioned before. As a result, the values of $n_s$ and $r$ obtained 
only for $N_\ast>60$ in the quadratic chaotic inflation model could be
realized even for $N_\ast=50$ - 60.
This clarifies a typical feature of the inflation induced in the model.
It is induced by the $c_2$ term in the potential.
As commented above, the trajectory cannot follow the potential minimum 
and the field suddenly rolls down towards the global minimum 
as long as the $c_2$ term is neglected.
Thus, in order to estimate these parameters including the value of 
$N(\chi_e)$, we need the analysis keeping the effect of $c_2$ terms 
as an indispensable one.\footnote{Here we confine our analysis to 
the case defined by $n=3$ and $m=1$, although other values of $n$ could 
give interesting results. Those detailed results will be presented 
elsewhere.} 

\begin{figure}[t]
\begin{center}
\begin{tabular}{c|cccc|cccc}\hline
&$c_1$ & $c_2$ & $\frac{\Lambda}{M_{\rm pl}}$ 
&$\frac{\varphi_\ast}{\sqrt{2} M_{\rm pl}}$ & $H_\ast$~(GeV) 
& $N_\ast$ & $n_s$ & $r$ 
\\ \hline\hline
A &$1.66\times10^{-6}$ & 0.7 & 0.04 &0.378 &  $0.871\times 10^{14}$ 
&59.0 & 0.971 & 0.107 \\ 
&$2.04 \times10^{-6}$ & 0.7 & 0.04 &0.371 & $0.921\times 10^{14}$ 
& 54.2 & 0.968 & 0.119 \\ 
&$2.42\times10^{-6}$ & 0.7 & 0.04 &0.366 &  $0.965\times 10^{14}$ 
& 49.1 & 0.965 & 0.131 \\ \hline
B& 0.257 & 6.0 & 0.002 &0.0512 &  $0.945\times 10^{14}$ 
&60.4 & 0.969 & 0.124 \\ 
&0.305 & 6.0 & 0.002 &0.0505 & $0.986\times 10^{14}$ 
& 55.0 & 0.966 & 0.136 \\ 
&0.364 & 6.0 & 0.002 &0.0498 &  $1.03\times 10^{14}$ 
& 50.0 & 0.962 & 0.149 \\ \hline
\end{tabular}
\end{center}

\vspace*{5mm}
{\footnotesize {\bf Table. 1}~~ Hubble parameter, spectral index and the 
tensor-to-scalar ratio for typical examples of four parameters of 
the model defined by $n=3$ and $m=1$. 
These model parameters are fixed to realize the observed value 
for the scalar perturbation amplitude $\Delta_{\cal R}^2$.} 
\end{figure}

In Table~1 we show typical examples which are calculated numerically
for different values for the model parameters $c_1$, $c_2$ and 
$\Lambda$. 
These examples suggest that sufficiently large e-foldings 
such as $N_\ast=50$ - 60 could be realized as long as $\Lambda \ll \varphi_\ast$ 
is satisfied even for the sub-Planckian inflaton value 
$\varphi_\ast< M_{\rm pl}$.\footnote{The values of $N_\ast$ in Table 1 are 
obtained through the direct numerical integration of Hubble parameters.
If we use the formula given in eq.~(\ref{efold}), the similar value can 
be obtained within a few percent difference from those. 
Both $n_s$ and $r$ are found to be not sensitive for such differences.} 
The Hubble parameter during the inflation takes 
values around $10^{14}$ GeV as shown in this table.
The predicted values of $n_s$ and $r$ are also listed in each case. 
In this calculation, we again use the solutions obtained 
from the field equations (\ref{evol}). 

In Fig.~3, we plot the predicted points in the $(n_s, r)$ plane 
for $N_\ast=50$ - 60 in the cases A and B given in Table 1. 
Although the model parameters $c_2$ and $\Lambda$ are required to take 
values in suitable regions to realize the observational data, the serious 
fine tuning of parameters seems not to be necessary.
As a reference, we plot the prediction of the $\phi^2$ chaotic inflation
as a dotted line in the same figure. Two crosses on it show the predicted 
points for $N_\ast=50$ and 60. 
The figure shows that this model could realize the $(n_s, r)$ points 
in the wider ranges which cannot be reached for $N_\ast=50$ - 60 
in the $\phi^2$ chaotic inflation scenario.  
It should be noted that that such points are realized even for the values of 
$N_\ast$ such as 50 - 60 by changing the model parameters suitably.
We find that the points of the case B somewhat shift from the line for the
quadratic chaotic model. This can be easily understood.
As found from eq.~(\ref{slow}), the $c_2$ term makes $\eta$ 
somewhat smaller than $\varepsilon$ compared with the $\phi^2$ 
chaotic inflation model where $\varepsilon=\eta$ is satisfied. 
Since this effect becomes larger for the parameters in the case A 
than the ones in the case B, the predicted points appear below the line for
the $\phi^2$ chaotic inflation model.
We find that the model is expected to predict $(n_s,r)$
in the region on or below the line predicted by the $\phi^2$ chaotic 
inflation model and also in the region where $n_s$ takes a larger value 
than the one predicted by the $\phi^2$ chaotic inflation model 
for the fixed $N_\ast$ value.   
These features show that the model could be an alternative 
interesting scenario to the simple $\phi^2$ chaotic inflation model. 
If both values of $n_s$ and $r$ could be constrained through the precise 
data obtained from the future CMB observations, the model could 
be tested in the near future. 

Finally, we should address the notorious $\eta$-problem in the model.
Although we could make higher order terms suppressed by the 
Planck scale in the potential ineffective, 
the $\eta$ problem still remains in the model. 
It may be formulated in two forms such that, 
(i) why $\frac{\Lambda}{M_{\rm pl}}$ in eq.~(\ref{model3}) is smaller 
than $O(1)$, and (ii) why $\tilde m_S^2$, $m_S^2$ and $\kappa_1\varphi^2$ 
are small enough in comparison with $H^2$.
As mentioned before, the first one is closely related to the UV completion
of the model, which fixes the exponential part of the potential. 
This could be solved only if the UV completion is found and 
its dynamics is clarified. It is beyond the present scope.
The second one might be also determined in the UV completed model.
In the present model, however, the values of $\tilde m_S^2$ and $m_S^2$ 
are related to other low energy physics, that is, the neutrino masses. 
This additional aspect might give a new physical meaning to the $\eta$ problem 
in this model. We might approach the problem based on this view point.

\begin{figure}[t]
\begin{center}
\epsfxsize=9cm
\leavevmode
\epsfbox{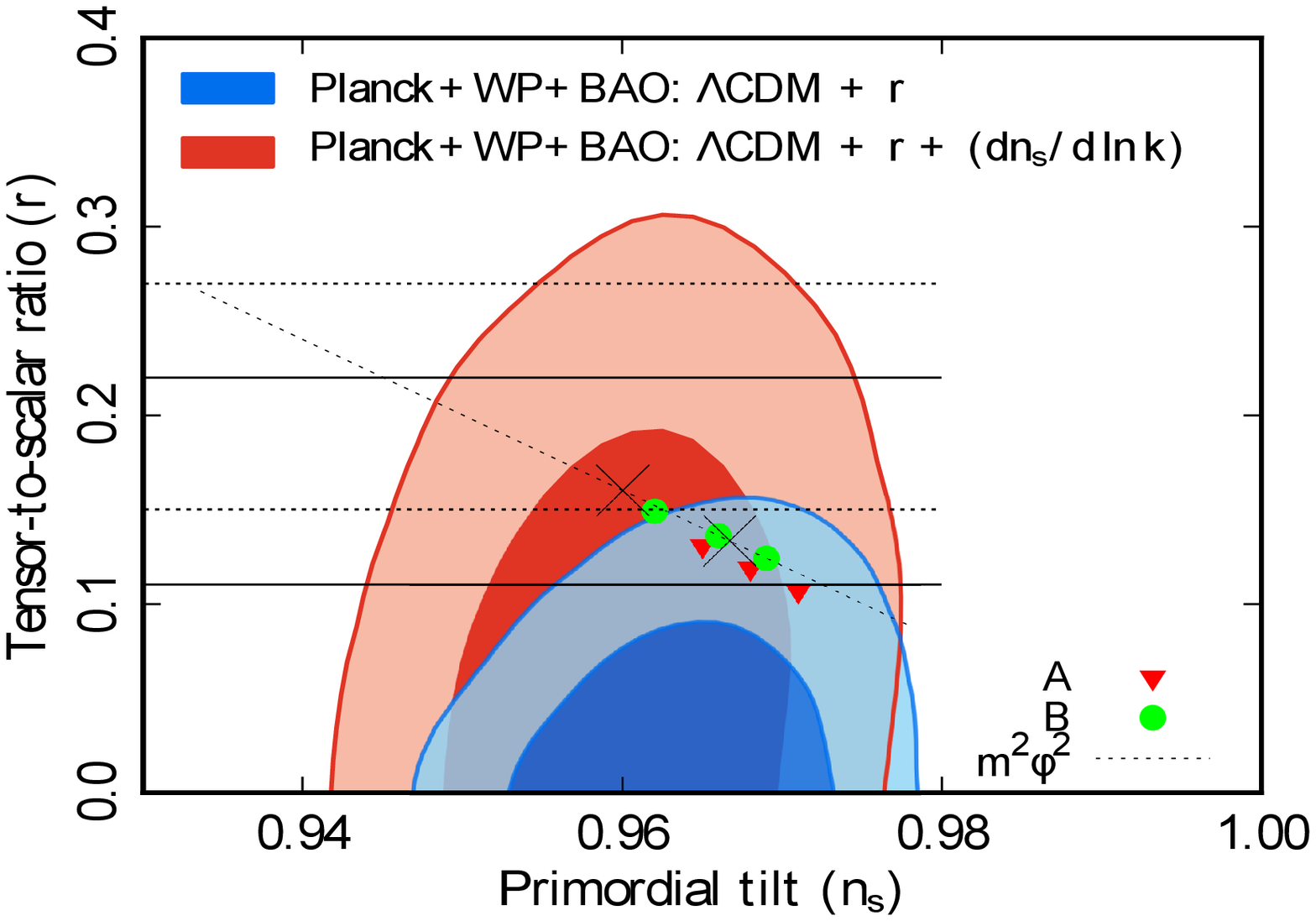}
\end{center}
\vspace*{-3mm}

{\footnotesize {\bf Fig.~3}~~Predicted values of $(n_s, r)$ for several 
parameter sets $(c_2,\frac{\Lambda}{M_{\rm pl}})$ given in Table 1. 
A dotted line represents the prediction by the quadratic chaotic inflation 
model and the crosses correspond to the points for $N_\ast=50$ and 60. 
Horizontal solid lines and dotted lines represent the Bicep2 
$1\sigma$ constraints with and without the foreground 
subtraction, respectively \cite{bicep2}.    
Contours given as Fig.~4 in Planck Collaboration XXII \cite{planck} are 
used here. Since the running of the spectral index $n_s$ is negligible in 
the present model, the blue contours should be compared with the 
predictions.}  
\end{figure}

\subsection{Reheating after the end of inflation}
The result shown in the previous part suggests that the model has favorable
features as an inflation scenario. In order for the model to be a realistic
one, it is required that the inflaton energy should be transferred
to radiation energy to reheat the Universe after the end of inflation.
It could be expected to occur if the aperiodic circular motion of 
the inflaton stops at a certain period and starts to behave as matter
through the oscillation around the global minimum of the potential.
In fact, such a behavior can be found to occur in the right panel of Fig.~2.
Since the kinetic energy of the fields becomes larger compared with 
the local potential barrier which gradually becomes smaller,
the field component $\varphi$ is expected to leave the local minimum and go 
over the potential toward the global minimum at a certain period. 

As reheating processes during the $\varphi_{1,2}$ oscillation, 
we have to consider both preheating due to 
the parametric resonance \cite{parares} through quartic interactions
of $S$ with $\phi$ and $\eta$ and also the perturbative decay due to an 
interaction term $\mu S\eta^\dagger\phi$ given in eq.~(\ref{model}).
Just after the end of inflation, the fields $\varphi_{1,2}$ 
start the oscillation around the global minimum with very large amplitude. 
Since the fields coupled with them have large effective masses and 
then it seems difficult for $\varphi_{1,2}$ to produce these particles. 
However, the particle production due to the parametric resonance 
is known to occur effectively even in such a situation. 
In this model the parametric resonance due to the scalar quartic 
couplings might realize rather high reheating temperature.

On the other hand, only preheating cannot transfer the inflaton energy 
to the radiation completely \cite{parares,reheat}. 
The decay of $\varphi_{1,2}$ induced through three scalars 
interaction such as $\frac{\mu}{\sqrt 2}\varphi_1\eta^\dagger\phi$ and 
$\frac{i\mu}{\sqrt 2}\varphi_2\eta^\dagger\phi$
can complete the energy transfer in this model.
The decay width for these processes is estimated as 
$\Gamma_{\varphi_i}=\frac{1}{8\pi}\frac{|\mu|^2}{\bar{m}_i}$ where
$\bar{m}_1^2=\tilde m_S^2+m_S^2$ and $\bar{m}_2^2=\tilde m_S^2-m_S^2$.
Since $\tilde m_S\gg m_S$ is assumed to be satisfied here, 
the reheating temperature realized through these processes could be 
estimated as \cite{reheat}
\begin{equation}
T_R\simeq 0.35~g_\ast^{-1/4}~|\mu|~\sqrt{\frac{M_{\rm pl}}{\tilde m_S}},
\label{tr}
\end{equation} 
where we use $g_\ast=116$ as the relativistic degrees of freedom 
in this model. Since both $\mu$ and $\tilde m_S$ are relevant to the 
neutrino mass generation as shown in the previous part, 
we should take account of the constraint from it.
The present inflation scenario also requires that eq.~(\ref{model3})
is the dominant potential of $S$ at the inflation era.
This brings about the additional constraints on $\tilde m_S$ as
\begin{equation}
\tilde m_S \ll \sqrt{c_1}
\left(\frac{\varphi_\ast}{M_{\rm pl}}\right)^{n-2}\varphi_\ast
\simeq 3.1\times10^{-4}\left(\frac{n}{N_\ast}\right)^{1/2}
\left(\frac{M_{\rm pl}}{\varphi_\ast}\right)^2\varphi_\ast,
\end{equation}
where eq.~(\ref{c1}) is used. If we apply $N_\ast=60$ and 
$\varphi_\ast\simeq 0.5M_{\rm pl}$ which are the typical values 
for the case $n=3$ in the previous part, the bound for $\tilde m_S$
can be obtained as $\tilde m_S \ll 3.4\times 10^{14}$ GeV.
Taking account of this constraint, we may estimate the reheating 
temperature through this process as
\begin{equation}
T_R\simeq 1.6\times 10^8\left(\frac{|\lambda_5|}{10^{-6}}\right)^{1/2}
\left(\frac{\tilde m_S}{m_S}\right)
\left(\frac{\tilde m_S}{10^6~{\rm GeV}}\right)^{1/2}~{\rm GeV}. 
\label{reheat}
\end{equation}
Here we also note that $|\lambda_5|$ should be smaller 
than $O(10^{-6})$ from the present bound of DM direct search 
since the lightest neutral component of $\eta$ is DM
and its mass is $\sim 1$~TeV \cite{ks}.
We find that the reheating temperature could be in a rather wide range
such as $10^5{\rm GeV}~{^<_\sim}~T_R~{^<_\sim}~10^{15}~{\rm GeV}$ depending on 
a value of $\tilde m_S$.
This temperature is high enough to produce thermal right-handed neutrinos
in the present model since the masses of right-handed neutrinos are assumed 
to be of $O(1)$ TeV. 
If the right-handed neutrino masses are sufficiently degenerate, 
the baryon number asymmetry could be generated through
the resonant leptogenesis as discussed in \cite{ks}. 
Right-handed neutrinos need not to be light but they could have large 
mass such as $O(10^9)$ GeV in a consistent way with this 
neutrino mass model \cite{ks}. 
Even in that case, eq.~(\ref{reheat}) shows that sufficient reheating 
temperature could be induced for leptogenesis to work well without 
the resonant effect. Anyway, the model could cause sufficient reheating 
temperature for the generation of baryon number asymmetry independently 
from the details of preheating in the model.

\section{Summary}
In this paper we have considered an extension of the radiative seesaw 
model with a complex singlet scalar to realize the inflation of 
the Universe keeping favorable features of the original model,
that is, the simultaneous explanation of the small neutrino masses,
the DM abundance and the baryon number asymmetry in the Universe.
This singlet scalar plays a crucial role not only 
in the exponential expansion of the Universe as an inflaton 
but also in the small neutrino mass generation.
In this scenario inflaton trajectory follows an aperiodic circle during the 
inflation.
This feature makes it possible that sub-Planckian values of the relevant 
field induce trans-Planckian changes of the inflaton value which is needed 
for the sufficient e-foldings. The model could be free from the serious 
problem caused by trans-Planckian field values.
However, the $\eta$ problem still remains. The UV completion of 
the model is expected to give a solution for it. 
 
We have also shown that the model has other interesting aspects 
as the inflation model. 
As a limiting situation, it gives the same formulas for the spectral 
index $n_s$ and the tensor-to-scalar ratio $r$ as ones of the $m^2\varphi^2$ 
type chaotic inflation, in which $r$ could take a large value. 
In more general cases, we have estimated them by solving numerically 
the field equations for the component fields of the singlet scalar. 
The tensor-to-scalar ratio could take large values in these cases also.
Both the spectral index $n_s$ and the tensor-to-scalar ratio $r$ could 
have values which are favorable from the recent CMB observations. 
If the precise data from the CMB observations are given in the near future, 
we could restrict the model parameters much more.
Since the roughly estimated reheating temperature tells us that 
it could be high enough for leptogenesis, the model seems 
to explain consistently the crucial problems in the SM including the baryon 
number asymmetry.
Although we cannot mention the origin of the specific 
potential at this stage, the features shown by the model seem to be interesting.
The model may deserve further study.    

\section*{Acknowledgements}
R.~ H.~ S.~ Budhi is supported by the Directorate General of Higher 
Education (DGHE) of Indonesia (Grant Number 1245/E4.4/K/2012). 
S.~K. is supported by Grant-in-Aid for JSPS fellows (Grant Number 26$\cdot$5862).
D.~S. is supported by JSPS Grant-in-Aid for Scientific
Research (C) (Grant Number 24540263) and MEXT Grant-in-Aid 
for Scientific Research on Innovative Areas (Grant Number 26104009).

\newpage
\bibliographystyle{unsrt}

\end{document}